# The layers that build up the notion of time

*Carlo Rovelli*

*Aix Marseille University, Université de Toulon, CNRS, CPT, 13288 Marseille, France.*
*Perimeter Institute, 31 Caroline Street North, Waterloo, Ontario, Canada, N2L 2Y5.*
*The Rotman Institute of Philosophy and the Department of Philosophy of Western Ontario University,*
*1151 Richmond St.N, London, Ontario, Canada, N6A 5B7.*

**1.**

Much confusion and disagreement around the notion of time is due to the fact that we commonly fail to recognize that we call "time" a variety of distinct notions, which are only partially related to one another.

The time of the researcher in quantum gravity (like myself), the time of the cosmologist, the time of scientist that studies black holes, the time of the engineering designing a GPS system, the time of biology, the time of the train-station master, the time of the historian, the time of the lover waiting for her love to arrive, the time of the old man thinking about his life and the time of the kid dreaming her future, are obviously somewhat related, but they are profoundly different.

For a particle physicist or a black-hole researcher, for instance, it is obvious that the time elapsed between two events depends on the path taken, while for the train-station master this is inconceivable and for the biologist it may be conceivable, but it remains totally irrelevant. The time in Newton equations does not distinguish past from future, the time of biology needs to. And so on.

None of these "times" is false. None of them is "illusory" or "not real". They are all, each of them, appropriate concepts for organizing phenomena at some scale, in some domain. But they are only valid on some scales, for different regimes, and in different contexts. Some connections between these times are obvious. For instance, the subtleties of relativistic time become negligible, irrelevant, when relative velocities are small. Other connections are more complicated, they have required centuries of scientific investigations to be sorted out. Some nodes of the overall story are not fully clear yet.

The common mistake, that I find repeated over and over again, is to take any one of these notions of time, good and appropriate in some domain, and mistakenly think that this particular one is universal, necessary, it is "what time really is", or "what we mean by time".

In other words, the common mistake is not in the fact that we understand time is this or that manner: it is to think that this or that manner are universal features of temporality. **To understand time, we have to break it apart**. We have to study its various layers [Fraser, 1975]: these have different properties, which in turn are sourced in approximations, in particular conditions [Rovelli 2018a].

I illustrate below a number of distinct notions of time and their differences; they are all relevant for describing the real world. What is important in what follows is to realize that many "obvious" properties of time are actually results of different kinds of approximations and idealizations.

**2.**

Let me start with a well-known example. Take two events. For instance, I clap my hands and call this event **A**. Then I clap my hands again and call this event **B**. We can say that between the event **A** and the event **B** some time has lapsed, and this elapsed time is a quantity $t_{AB}$ that can be characterized as the reading of a (good) clock. This is a definition of time that can make many physicists, sport arbiters, and engineers, happy. How universal is it?

In our everyday experience, it largely suffices. But any good relativist knows well that this does not work in general. The reason is that, if they are sufficiently precise, two identical clocks measure different values $t_{AB}$ if they move differently between the two events **A** and **B**. Any clock that moves (fast enough) away from **A** and then comes back (fast enough) to **B** measures a *shorter* time than a stationary clock. So the time lapse between **A** and **B** is in fact arbitrary short, if the clock measuring it is moved fast enough. This is a fact.

We may try to correct the previous definition by stating that the "true" time lapse between the event **A** and the event **B** is the time measured by a stationary clock: a clock *that remains in my hand* and does not move. This, we may think, is the *longest* possible time span between **A** and **B**.

But this would be wrong, again. The clock that measures the *longest* time span between **A** (the first clap of my hands) and **B** (the second clap of my hands) is actually not the clock that remains near my hands. It is a clock that I throw upward at the moment of **A** and catch back when it falls down precisely at **B**. (The reason is that this clock is in free fall during the flight, hence follows a geodesic, which can be shown to maximizes the *proper time* between the extremes of the flight). We could define time in this way, but it follows that if we do so we cannot synchronize clocks over a region. If I clap three times, the total time is not the sum of the two intermediate times.

There is no common clock time: each clock measures its own 'proper' time along its path, and the proper times do not match on arrival. The time of a relativist is definitely quite different from the universal common clock time of our daily experience.

This is just how Nature happens to work. There is nothing wrong in the idea of a **common universal time**, but this idea only works within an approximation where we disregard relativistic effects. **General relativistic time** is not unique and not universal.

### 3.

The same is true with regard to cosmological time. Cosmologists use the notion of "**cosmological time**", which is roughly defined as the time that a clock sitting inside a galaxy would have measured it had started at the Big Bang. But this notion of time is an approximation, because when two galaxies merge (as our Milky Way and Andromeda are headed to do in the future) they have in general different proper times from the Big Bang. Which of the two has the "right" time? The question is obviously meaningless. Cosmological time is only defined under the (false) assumption of exact homogeneity. Beyond this approximation, cosmological time does not correspond to clock time.

Even within the homogeneity idealization, the simultaneity surfaces defined by cosmological time are not simultaneity surfaces in the sense of special relativity, as can be easily shown. Cosmological time is a fully conventional arbitrary labeling of spacetime events.

Neither our common non-relativistic time, nor cosmological time, not full relativistic times are "true times". They are related but distinct temporal notions, with different properties. The unicity of our common-sense notion of time is the result of an approximation, like the apparent flatness of the surface of the Earth.

### 4.

A good example of the misunderstanding caused by the failure to appreciate that the notion of time is complex and not monolithic is the erroneous argument that Einstein's relativity implies a static four-dimensional 'block' universe [Putnam 1967].

The argument goes as follows. Given two events **A** and **B**, with **B** in the past of **A**, the theory of relativity states that it is possible to find a third event **C** that is "simultaneous" to **A** with respect to some observer but also "simultaneous" to **B** with respect to some other observer. Since two simultaneous events are said to be both "real now" in common language, it follows that all the events of the universe are all "real now". Hence the universe is a four dimensional "block universe", which is "without becoming".

The argument is wrong because it is based on the mistaken assumption that the only manner time can function is as in our naïve non-relativistic intuition, where simultaneity can be defined in a transitive and observer independent manner. Einstein's definition of simultaneity is relational and is not transitive. This misunderstanding leads to **the false alternative between Presentism and Eternalism** [Rovelli 2019].

Presentism is the picture of reality based on the idea that there is a common objective "now" all over the universe. This idea is based on the commonsense impression that what we see around us is the is the common and objective "now". The discoveries of the physics of the last century have conclusively shown that this view is undefendable as the idea that the Earth is the center of the Universe. The commonsense impression that what we see a shared "now" is based on the fact that our senses are imprecise and we do not resolve the light traveling time. It is much like the fact that around us we see that the Earth is flat.

Eternalism is the idea that there is no temporal becoming, and our sense of temporal becoming is illusory. What the universe truly is, according to the Eternalist, is a static four-dimensional block, where future events are "real now" as present events are. This too is an absurd position, because the four-dimensional spacetime of relativistic theories is a manifold of events, namely of "happenings", and there is nothing "static" about them. There is nothing in relativity that motivates us to say that a future event is "real now", in any reasonable sense. **General relativity is about happenings, not about anything static**.

The source of the confusion is the mistaken idea that either time is exactly as in our naïve intuition, or it does not exist, it is illusory. This leads to the mistaken idea that Presentism and Eternalism are the sole alternatives. They are not. Relativity offers a third possibility, where things happen, but there is no global present. A physicist working with relativity easily develops the proper intuition to deal with a local becoming that is not organized in a single global shift from a universal now to the next [Rovelli 2019]. It is a notion of time that lacks the global character.

**5.**

Before Newton, in the long intellectual tradition that goes from Aristotle [Aristotle] to Descartes, as well as in common parlance, "time" was mostly understood as a counting of events: day, night, day, night… It is Newton who introduced the idea that time passes "by itself" irrespectively from the events happening. (In his famous defense of the relational notions of space, against Newtonian's absolute space, Leibniz was largely articulating (also with new arguments) notions of space and time that were traditional, not novel.) A few centuries later, Newtonian time looks "natural" to us.

Detaching the idea of an autonomously flowing time from the actual events of the world was instrumental for Newton to build the conceptual structure of mechanics, and proved a very effective move.

This implies that there are two distinct notions of times that are often at play (Newton clearly distinguishes the two in the *Principia* [Newton]). The first is the old notion of time as the simple *relative* succession of the events. The second is the Newtonian idea of a metric (measurable) time flowing *by itself*.

The difference is substantial. For instance, if truly nothing happens, Newtonian time continues to pass; while it is meaningless to say that the counting of happenings continues if nothing happens.

The traditional notion is *relational*: time is a relation between events. No events happening, no time. While the novel notion of time introduced by Newton refers to a (peculiar) *entity*, which exists by itself.

Once again, neither of the two is right or wrong: **relational time and Newtonian time** are just two notions playing a role in the organization of the phenomena.

Newtonian time definitely turned out to be far more effective than relational time in science for a few centuries. But Einstein's revolution changed the game in an unexpected manner, which in a sense lead to a compromise between the two previous notions of time. Newtonian time was first brought together with Newtonian space (another peculiar *entity* introduced by Newton) to form *spacetime*. Then, this spacetime itself was identified by Einstein as a physical field, the gravitational field. As such, it is definitely an entity, but not a *peculiar* entity anymore. It is a rather mundane entity: a field like the other fields that modern physics uses to describe the world. Notice that even the causal structure of spacetime is just a property of a field, the gravitational field.

Thus, Newtonian time is reduced to an aspect of a field, an ensemble of events itself, after all. The traditional relational notion of time, on the other hand, remains useful, of course. We can still count events in a general relativistic context and call "time" their relations, which are well accounted for by the mathematics of general relativity. We can still count day, night, day, night… and call this time, also in a general relativistic Solar System.

Again, we have a multiplicity of notions of time, and we understand their relations: how some emerge within approximations from others in special regimes.

**6.**

There is no consensus on a **quantum theory of gravity** yet, and no direct empirical support for any of the current tentative theories. But there are tentative theories of quantum gravity, that coherently and consistently merge quantum theory and general relativity. Among the best is *loop quantum gravity*, with which I work [Rovelli 2004].

Loop quantum gravity describes the quantum behavior of the gravitational field, namely of (relativistic) spacetime. It describes for instance the possibility that between two events **A** and **B** even the time measured by *the same* clock could be in quantum superposition of two different values.

Recently, the possibility of measuring this effect (predicted by all current quantum gravity tentative theories) in the laboratory is been considered by some laboratories [Bose 2017, Marletto 2017].

In loop quantum gravity, the relational notion of time, namely the possibility of counting series of events, remains in place, but the Newtonian time, like the proper times of general relativity are replaced by quantum variables, which, like all quantum variables, can be discrete, have a probabilistic dynamics and be in superposition [Rovelli 2018b].

Hence **the basic equations of quantum gravity are very likely not to include a time variable** [De Witt 1967]. The basic equations of loop quantum gravity, indeed, don't have a time variable [Rovelli 2004].

This, however, is no big deal, because classical general relativity can be equally formulated without using a time variable, for instance in its Hamilton-Jacobi formulation. There is nothing mysterious in the absence of a time variable, in spite of the discussion that his fact has generated and unfortunately still generates. The reason for the absence of a time variable is simply that in general relativistic physics, as mentioned above, there is no preferred time variable, no preferred interval between **A** and **B**. Hence, it is rather obvious that the theory can be formulated without a preferred time variable. The theory gives the **relative evolution** of physical variables with respect to one another. For instance, it gives the evolution of local variables with respect to local clocks times, and the **relative evolution** of clock times with respect to one another.

Consider for instance the example described above, where I keep one clock in my hand, throw the second upward and catch it when it falls back. It is a fact that the time between the launch and catch measured by the two clocks is different. Which of the two measures the *true* elapsed time? Is the theory describing how the flying clock evolves in the *true* time defined by the stationary clock, or vice-versa? The question is meaningless: the theory tells us how the clocks change one with respect to the other.

The conceptual structure of the theory is clear and well defined without any need to single out a preferred variable and call it the real time [Rovelli 2011].

Loop quantum gravity allows us to (tentatively) compute what happened around the Big Bang, in the interior of black holes or in other situation where spacetime is not classical, in a comprehensible and predictive (although not yet empirical supported) way, without any need of having a specific time variable, or a preferred notion of time, besides the ancient relational idea that we can call time any counting of events.

**7.**

A striking feature of the entire structure of elementary physics (classical mechanics, relativistic field theory, quantum mechanics [see DiBiagio 2020], quantum field theory, the Standard Model, general relativity…) is the fact that the basic equations to do not distinguish the past from the future (provided we also swap parity and charge, namely swap the names "left" with "right" and "positive" with "negative" charge) [Price 1997].

This is in striking contrast with the manifest irreversibility of most phenomena around us, that makes the past direction of local time dramatically different from its future direction.

All observable distinctions between past and future can be traced to a fact: the tendency to equilibrate towards the future, but not towards the past [Myrvold 2021].

Contrary to what the study of the statistical mechanics of gases suggests, in the real-world equilibration is generally very slow (the universe is still very far from equilibrium). Many realistic thermalization times are huge (in fact, cosmological). The Sun is hot, the Earth cool, after billions of years of slow equilibration.

While equilibration towards the future is intuitively comprehensible, the lack of equilibration towards the past is puzzling. This is usually denoted as the puzzle of the past low entropy, or the statement of the "past hypothesis". In the following paragraph I mention some current ideas on the possible ground of this macroscopic asymmetry: here I discuss what it implies about time.

The key to understand what this macroscopic asymmetry implies about time, it to realize that the difference between the past and the future pertains to the **macroscopic** description of phenomena only. **There is no way of even naming it, in terms of microphysics alone**. While some macroscopic histories (breaking eggs) are more commonly observed than their time reversals (eggs recombining), every single micro history is equally peculiar and equally unlikely to be found in the universe both in its ahead and in its time reversed version. (The reason is that no single microhistory is *ever* commonly observed.). It is only when we patch micro-histories in groups (when we coarse grain) that we begin to see irreversibility: in the past, the actual micro history of the world belongs to a macro state with an intact egg; in the future it belongs to a much wider macro state of broken eggs. **Irreversibility is a property of the macro-variables, not the micro-history**.

It is important to stress that the time orientation of *all* irreversible phenomena is grounded in this basic arrow of time. Irreversible phenomena are all macroscopic and statistical in nature. That is, they all exist thanks to a microphysics with many degrees of freedom where energy dissipates. Among these are the fact that we have traces of the past (and not corresponding traces of the future) [Rovelli 2020a], agency [Rovelli 2020b] and time travel [Lewis 1993]. The fact that all arrows of time depend on the entropic one is important and is often misunderstood.

An important consequence of this fact is that since phenomena like thinking, being conscious, and similar are all irreversible, they are necessarily macroscopic. They cannot be elementary.

The key point here is that time orientation is not a necessary property of some "elementary" basic notion of time. It is only a feature of the approximate notion of time we use in coarse grained "approximate" accounts of the natural phenomena. **The orientation of time is ubiquitous but is not fundamental**.

The time orientation of our thinking, of our living, of biology, of evolution, of the notion of causation that we employ, the abundance of traces of the past that has no equivalent in the future… all these phenomena are grounded in the macroscopic approximation. They are all (in a wide acceptation of the term) of entropic origin. The time of the microphysics has no orientation. There is no contradiction between this fact and the fact that we have any reason to believe that everything we observe supervenes on the microphysics.

This is strongly counterintuitive, but it is an inescapable consequence of what we have learned so far about the natural world. The very specific "temporal" features of the time variable has to be studied with statistic, thermodynamics [Rovelli 1993] and perhaps the lack of full information implied by quantum theory [Connes & Rovelli 1994], not in any *a priori* intuition about time flow. Our intuitions have developed for our macro world, no surprise they are misleading

**8.**

The existence of the entropy gradient in the macroscopic account of our universe remains puzzling. It can be taken as a contingent fact or as a law ("the past hypothesis") [Albert 2000]. It might also be a perspectival phenomenon. **The possibility that the arrow of time is perspectival** [Rovelli 2016] is a speculative but intriguing possibility.

It is based on the fact that any macroscopic account of phenomena is determined a choice of macroscopic variables. This choice is not arbitrary: it is dictated by the available physical interactions with the system. Hence any macroscopic account is fundamentally relational: it is relative to a **second** physical system that interacts with the first via an interaction that involves only a (relatively) small number of varaibles: these variables are the macroscopic variables.

This observation opens up the intriguing possibility that the asymmetry of the arrow of time is not due to fact that the past microstate is by itself non-generic. Rather, it is due to the coarse graining defining the macro-physics. This is determined by the physical way we interact with the rest of the universe. Hence it may be that what is special is not the micro-history of the world, but our being part of a subset of the world that interacts with rest via relatively few macroscopic variables, *which define the peculiar coarse graining with respect to which there is an entropy gradient*.

If this is correct, the distinction between past and future could be a majestic phenomenon, but a perspectival one: like the apparent rotation of the universe around us. The rotation of the sky we observe around us is a perspectival phenomenon, due to the fact that we live on a spinning planet. It is real but perspectival phenomenon.

To understand time, as to understand all the phenomena we witness, we have to take into account the perspective from which we witness them [Ismael 2007].

**9.**

**Experiential time** is a far more complex phenomenon that physical time. I believe that a lot of confusion about time is based on the fact that we have a a rich temporal experience and we erroneously pretend to project it all over elementary physics, where it does not belong [Husserl 1928]. In this, we are as primitive as the ancient Greeks when they tried to understand the dynamics of the material world in terms of the "basic forces" of "love" and "hate".

Experiential time is generated by the way our brain works. A specific activity of the brain is to store information about its past events. This is possible precisely because the brain is a macroscopic object functioning in an environment far away from equilibrium, with a constant income of free energy (we live nearby an infernal furnace burning at 6000 degrees, the Sun). Hence the functioning of the brain does not make up a time arrow: it gets it from the macroscopic physics that defines it and the entropic time arrow of the environment.

The brain stores information in the form of memories and constantly retrieves and utilizes some of these memories. This has the effect that in a sense our experiential present is impregnated by aspects of the past.

Furthermore, recent neuroscience suggests that one of the main activities of the brain (if not *the* main activity) is to constantly compute possible futures trying to anticipate them [Buonomanno 2017]. Hence we equally live with vivid awareness of some future (possible) events.

The result of all this is that our experience includes a sort of temporal **window** formed by the set of events included in memories and anticipations. This is the window we actually call "time". (See the notion of *Lichtung*, in Heidegger 1950).

We presumably have this basic structure in common at least with mammals. Our species, on the other hand, is probably characterized by an increased span of this window, thanks to collective memory (and science) that give us access to larger chunks of past than other mammals, and a greatly increased capacity of future planning, compared to other mammals, that give us a wider, and perhaps richer (or poorer?) sense of the future.

In any case, we literally do not experience just the present, but also the past and the future. This experience of time depends on the fact that we have a brain. The beautiful clock in my living room keeps time much better than me, but it has no comparable experience of the past and the future in any remotely connected sense. It does not remember yesterday and is not waiting for tomorrow. Our experiential time is grounded in our complexity. The sense of the speed at which times flows, the nostalgia of the past, the anticipation of the future, all the rich phenomenology of our experience of time obviously are grounded in physics, but are not universal aspects of physical temporality.

Yet, it is hard to make abstraction from them. It is hard to renounce the intuition that the Big Bang, 14 billion year ago is "very far", while what happened 3 seconds ago is "just now", and to focus on the fact that the sense of closeness or distance is an effect of the specific capacity of our memory. The perceived speed at which time passes has obviously far more to do with the working time of our neurons than to anything pertaining to mere physics.

To understand time, we have to learn to separate our intuition about time, which is grounded in this rich experiential time, from the concepts of time that turn out to be needed to describe the events of nature. I think that too often we fail on this.

**10.**

Here is a brief table listing some notions of time with their properties

| **Time notion** | **Some typical properties** | **Ground** |
| --- | --- | --- |
| *Experiential time* | Feels as having a flowing speed. | Brain structure |
| *Irreversible time* | Oriented. Past differs from future. | Disregard microscopic degrees of freedom |
| *Newtonian time* | Unicity. Single now in the universe. | Disregard relativistic phenomena |
| *Special relativistic time* | Independent from actual events happening. | Disregard relativistic gravity |
| *Cosmological time* | Global. | Disregard inhomogeneity |
| *Proper time in general relativity* | Measured by local clocks. | Disregard quantum phenomena |
| *Independent evolution variable in quantum gravity* | Arbitrary. | Not needed to describe happenings |
| *Relational time* | Always available. | Arbitrary counting of happenings |

**10.**

Experiential time is not only heavily infiltrating and (mis-)guiding our intuition, when we try to understand the temporal structure of reality, but is also heavily emotionally charged.

Two of the most influential investigators of the nature of time, who are perhaps at the farthest extremes of the cultural spectrum, are Hans Reichenbach [Reichenbach 1958] and Martin Heidegger [Heidegger 2010]. Quite surprisingly both of them point out that much of philosophy —and other human enterprises— can be read as efforts to escape the anxiety that time causes.

I suspect that this anxiety is not just an emotional disturbance that fogs our efforts to get intellectual clarity about the notion of time. It is actually the other way around: time for us is largely precisely this anxiety, the emotional underpinning that drives the constant process of utilizing memories to build our future or protect us from it.

Separating these entrenched intuitive aspects of *time-for-us* form the temporal notions we use to organize and understand the physical world does not mean that we have to choose one of the two times as the true one, as opposite intellectual traditions unfortunately do. Indeed, I do not think that it makes much sense to say that experiential time is the *authentic* one, as some contemporary schools pretend, while natural time is constructed. Nor I think it makes much sense to say that physical time (which one?) is real, and what we experience is illusory, as other contemporary schools pretend.

To get clarity about the complex structure of the natural world, of which we are part, we have to distinguish the multiple layers that build up the complex phenomenology of time. We have to get some clarity about the multiplicity of structures that we negligently and generically call "time".